\documentclass[runningheads]{llncs}
\usepackage{amsmath,amssymb,amsfonts}
\usepackage{algorithm}
\usepackage{algpseudocode}
\usepackage{graphicx}
\usepackage{booktabs}
\usepackage{multirow}
\usepackage{xcolor}
\usepackage{tikz}
\usetikzlibrary{arrows.meta,positioning,fit,backgrounds,calc}
\usepackage[hidelinks]{hyperref}

\begin{document}

\title{Real-Validated UAV Audition Under Rotor Ego-Noise for Low-False-Alarm Human Detection}
\titlerunning{Real-Validated UAV Audition Under Rotor Ego-Noise}
\author{Junhao Wei\inst{1}\orcidID{0009-0006-0553-2032} \and
Haochen Li\inst{1}\orcidID{0009-0000-8213-5854} \and
Dexing Yao\inst{1}\orcidID{0009-0007-6267-8967} \and
Yanxiao Li\inst{1}\orcidID{0009-0008-3389-1619} \and
Yifu Zhao\inst{1}\orcidID{0009-0004-2363-9269} \and
Baili Lu\inst{1}\orcidID{0009-0000-3814-2918} \and
Zhenhong Peng\inst{2} \and
Ngai Cheong\inst{1}\orcidID{0000-0002-5599-4300} \and
Xu Yang\inst{1}\orcidID{0000-0002-7037-3609} \and
Yapeng Wang\inst{1}\orcidID{0000-0002-1085-5091}
\thanks{Corresponding author: \email{yapengwang@mpu.edu.mo}}}
\authorrunning{Junhao Wei et al.}
\institute{Faculty of Applied Sciences, Macao Polytechnic University, Macao, 999078, China
\and
College of Animal Science and Technology, Zhongkai University of Agriculture and Engineering, Guangzhou, 510225, China}
\maketitle

\begin{abstract}
Detecting human acoustic cues---speech, cries, coughs---from a microphone mounted on an
unmanned aerial vehicle (UAV) could help acoustic search and rescue (SAR), but rotor
ego-noise buries the human signal at extremely low signal-to-noise ratios (SNR), often below
$-30$~dB, and large openly labelled airborne corpora are scarce. This data constraint has pushed
the field toward synthetic training and evaluation, while the relationship between synthetic
scores and real UAV behaviour remains weakly characterized. We build a real-validated benchmark
for UAV human-audible-presence (HAP) detection: models are trained on a reproducible synthetic
mixture pipeline assembled from public audio, but selected and reported on real DroneAudioSet
using an operating-level audibility filter, a recording-grouped real Dev/Test split, and group
bootstrap confidence intervals that resample recordings rather than windows. Two findings are
robust. First, synthetic accuracy is a poor, non-monotonic proxy for real transfer: a
from-scratch SE-ResNet reaches synthetic AUROC $0.743$ yet collapses to real-operating AUROC
$0.536$, whereas frozen audio foundation models and their adapters transfer far better
($0.72$--$0.79$ real AUROC). Second, once selection is done on real operating data, a
domain-regularized BEATs adapter chosen by our pre-registered Real-Dev rule gives the best
low-false-alarm operating point among the candidates (recall at $5\%$ FPR of $0.559$ on the locked
test, versus $0.483$ for a vanilla adapter). This method advantage is not statistically
significant under group bootstrap, because adapter variants cluster within noise on the limited
real recordings, and extreme-SNR positives remain near the detection floor for all methods. The
main contribution is therefore the benchmark and evaluation protocol, together with evidence that
real, group-level validation is necessary for honest conclusions in UAV audition. Code, recipes,
and splits will be released.

\end{abstract}

\section{Introduction}\label{sec:intro}
Small UAVs are increasingly used in search-and-rescue (SAR), where a rotorcraft can rapidly
cover terrain that is unsafe or slow for ground teams. Related autonomy research spans
adaptive particle-swarm path planning~\cite{wei2024adaptive}, three-dimensional planning with
GeoSSA and CICDWOA~\cite{wei2026geossa,wei2026cicdwoa}, and online operator selection for
inspection routing~\cite{wei2026labhh}. SAGA and KIO-planner investigate attention-based
navigation from depth observations~\cite{wei2026saga,yao2026kio}. ARIES-Mission2 combines visual
and language inputs with route optimization for aerial missions~\cite{wei2026ariesmission2}.
Movable antennas mounted on UAVs support integrated sensing and communication~\cite{long2026antenna}.
These advances also motivate the study of acoustic sensing for UAV missions.

Beyond cameras, an on-board microphone
could help localize survivors by their voices, cries, coughs, or other human acoustic cues even
when they are visually occluded by rubble, foliage, or smoke. This capability, which we refer to
as UAV audition, is dominated by the platform's own rotor ego-noise: the propellers generate
intense, strongly harmonic, time-varying sound that masks human cues at SNRs routinely below
$-30$~dB. Reliable detection of a human acoustic presence under such conditions is the problem we
study.

Progress is limited by both data and modelling constraints. Real airborne recordings with
controlled, labelled human sources are scarce and heavy; the most complete public resource,
DroneAudioSet~\cite{droneaudioset}, is tens of gigabytes and covers a limited set of platforms
and rooms, making broad supervised training difficult for many groups. At the same time, prior UAV
acoustic work is dominated by detect-the-drone systems that classify whether a drone is present
from ground microphones~\cite{alemadi}, rather than by drone-embedded audition that hears through
the platform's own ego-noise to detect a human. Audio foundation models such as PANNs~\cite{panns},
AST~\cite{ast}, and BEATs~\cite{beats} provide strong transferable representations, but their
behaviour under UAV rotor noise and the best way to adapt them remain under-tested on real
airborne recordings.

Because real airborne data is scarce, much of the plausible progress on this problem is made on
synthetic mixtures and then validated on those same synthetic distributions. This creates a proxy
risk: the model that separates synthetic mixtures best may not be the model that transfers to a
real UAV microphone. We address this risk by training on a fully reproducible synthetic pipeline
built from public audio, but selecting models and reporting final operating metrics only on real
DroneAudioSet recordings. The protocol uses a metadata-defined audibility filter, a
recording-grouped Dev/Test split, and group-level bootstrap intervals, so correlated windows from
the same recording do not masquerade as independent evidence. Within this protocol we evaluate
parameter-efficient adaptation of frozen audio foundation models, including variants conditioned
on rotor harmonic structure and regularized against synthetic nuisance domains.

The contributions of this paper are as follows: we provide a real-validated benchmark, quantify
the synthetic-to-real proxy gap, and evaluate a lightweight foundation-adapter family under a
low-false-alarm operating metric.
\begin{enumerate}
\item We introduce a reproducible UAV human-audible-presence benchmark that combines public
synthetic mixtures with a real DroneAudioSet grouped split. The real split is
grouped by recording, uses a metadata-only operating audibility threshold, and reports group
bootstrap confidence intervals.
\item We show a strong, non-monotonic proxy mismatch between synthetic and real performance. A
from-scratch CNN appears competitive synthetically but collapses on real ego-noise, while frozen
foundation encoders and adapters transfer much more reliably.
\item We analyze a BEATs adapter family that trains only about $1\%$ of the parameters. The
Real-Dev-selected EgoRAP-DA configuration gives the best locked-test low-false-alarm recall among
the candidates, while its gain over a vanilla adapter is reported transparently as
non-significant under paired group bootstrap.
\end{enumerate}

\section{Related Work}\label{sec:related}
\subsection{UAV acoustic perception}
Most UAV acoustic research targets drone detection and identification from ground sensors, using
log-mel or MFCC features with CNNs, ResNets, or attention backbones~\cite{alemadi}, as well as
multi-sensor or array-based localization~\cite{dregon,uavirbase}. The complementary
problem of drone-embedded audition, in which the microphone is mounted on the rotorcraft itself,
has been studied for source localization and enhancement under array processing~\cite{dregon} and
for SAR human detection with DroneAudioSet~\cite{droneaudioset}. Our work focuses on
human-audible-presence detection from a single UAV-mounted channel and evaluates the models under
a real grouped protocol rather than only on synthetic mixtures.

\subsection{Audio representations and foundation models}
Task-specific human acoustic representations include Graph-LSTM models for speech emotion
recognition~\cite{li2023graphlstm} and the fusion of convolutional, residual, and ECAPA-TDNN
components for speaker recognition in CRET~\cite{li2025cret}.
Large pretrained audio encoders such as PANNs, AST, and BEATs provide transferable
representations for audio tasks~\cite{panns,ast,beats}.
Related Portuguese speech recognition studies examine transfer learning with small datasets~\cite{wang2022smalldata}
and the feasibility of fine-tuning large end-to-end models~\cite{li2025portuguese}.
These studies provide context for representation learning under data constraints; transfer to
rotor-masked human-presence detection requires separate evaluation.
Parameter-efficient adaptation with small heads, adapters, or prompts~\cite{houlsby,lora} keeps
the backbone frozen while adding task-specific trainable parameters. EgoRAP extends this adapter
idea with a signal-derived rotor descriptor and rotor-modulated pooling, but the experiments
treat rotor conditioning as a candidate inductive bias to be validated on real data rather than
as an assumed source of improvement.

\subsection{Robustness and domain generalization}
Low-SNR audio detection is highly sensitive to source, device, and acoustic-domain shift.
Data augmentation has also been studied with ECAPA-TDNN for speaker recognition,
including changes to training-data diversity and availability~\cite{li2023augmentation}.
SpecAugment-style perturbations~\cite{specaugment}, domain-adversarial training~\cite{dann}, and
supervised contrastive learning~\cite{supcon} can encourage nuisance-invariant features, but their
effect can be misread when validation remains synthetic. Our results show that a regularizer that
reduces synthetic AUROC can improve real low-FPR recall, which makes real operating validation a
necessary part of the model-selection loop.

\section{Problem Formulation}\label{sec:problem}
Let $x(t)$ be a single-channel $T$-second waveform captured by a UAV-mounted microphone. The
primary task is binary human-audible-presence (HAP) detection: predict $y\in\{0,1\}$, where
$y{=}1$ iff a human acoustic cue usable for SAR is present---speech, human vocal non-speech
(scream, cry, cough, laugh), or human-produced non-vocal sound (clap, footstep)---and $y{=}0$ for
rotor-only, non-human environmental, or silent input. As an auxiliary signal we also predict a
four-way source class $c\in\{\textsf{none},\textsf{speech},\textsf{vocal-nonspeech},
\textsf{produced-nonvocal}\}$; the auxiliary head is used only for training regularization and
analysis, not as the reported metric.

We characterize synthetic difficulty by the SNR between the human source and the rotor ego-noise.
Because low-SNR positives are often below the detection floor, the synthetic analysis distinguishes
an operating range $[-45,-5]$~dB, an extreme regime $[-55,-45]$~dB, and the full range
$[-55,-5]$~dB. The headline results are reported on the real operating regime
(Sec.~\ref{sec:data}), not on any synthetic SNR band.

\section{Sim-to-Real Data Protocol}\label{sec:data}
\subsection{Public training sources}
The synthetic training pipeline uses only public audio sources. Rotor ego-noise comes from the
DroneAudioDataset (Bebop, Mambo, and generic ``yes-drone'' recordings)~\cite{alemadi}; speech
comes from LibriSpeech \texttt{dev/test-clean}~\cite{librispeech}; human vocal non-speech,
human-produced non-vocal sounds, and environmental negatives come from ESC-50~\cite{esc50}. A
held-out real slice of DroneAudioSet~\cite{droneaudioset} is used only for real OOD validation
and testing.

\subsection{Online synthetic mixing}
Each 4~s, 16~kHz training clip is generated on the fly as
\begin{equation}
x = a_h\,s_{\text{human}} + s_{\text{drone}} + a_b\,s_{\text{bg}},
\end{equation}
where the human gain $a_h$ realizes a target SNR sampled uniformly within one of five buckets
spanning $[-55,-5]$~dB. Positives ($50\%$) mix a drone with a human source
(speech~$60\%$/vocal-nonspeech~$30\%$/produced-nonvocal~$10\%$); negatives ($50\%$) are
drone-only ($40\%$), drone+environmental ($50\%$), or environmental-only ($10\%$). Six acoustic
augmentation profiles, covering microphone EQ, rotor-RPM amplitude modulation, reverberation, and
wind-like low-frequency noise, define an acoustic nuisance domain, and SpecAugment is applied to
the log-mel view.

\subsection{Synthetic stress folds}
The synthetic side includes source-disjoint splits whose purpose is analysis rather than final
model selection. Speakers and ESC folds are never shared across the relevant train/evaluation
sets. We use M0 for in-distribution evaluation, M1 for leave-one-drone-noise-out evaluation,
M2 for extreme-SNR transfer after training above $-45$~dB, and M3 for unseen speakers and an
unseen ESC fold.

\subsection{Real operating regime and grouped testing}
Many positive DroneAudioSet recordings contain distant or quiet sources that are physically
near-inaudible under rotor ego-noise. We therefore define the evaluable operating regime from
metadata alone. The received-level proxy is
\begin{equation}
L_{\text{recv}} = L_{\text{src}} - 20\log_{10}\!\big(\max(d,1)\big),
\end{equation}
where $L_{\text{src}}$ is the annotated source level and $d$ is distance. Real operating
evaluation includes audible positives with $L_{\text{recv}}\!\ge\!\tau$ versus all negatives. We
set $\tau{=}75$~dB from the metadata distribution, report a 70/75/80~dB sensitivity sweep, and
never tune the threshold on model outputs.

Real recordings are cut into $4$~s windows with a $1$~s hop and partitioned at the recording
level. The group key is the recording condition, including room, drone, microphone, distance, and
throttle, and every window inherits its recording's split. A stratified group split assigns
$40\%$ of recordings to Real-Dev and $60\%$ to Real-Test. From $264$ recordings this yields, at
$\tau{=}75$~dB, $1{,}109$ audible-positive windows over $38$ operating recordings in Dev and
$1{,}711$ over $58$ operating recordings in Test. All model and operating-point selection uses
Real-Dev only; Real-Test is scored once after the configuration is fixed. Every real confidence
interval is a group bootstrap over recordings ($2000$ resamples), and every method comparison
uses a paired group bootstrap, because window-level intervals would overstate precision.

\section{Adaptation Method}\label{sec:method}
\begin{figure}[t]
\centering
\resizebox{\textwidth}{!}{%
\begin{tikzpicture}[
  node distance=6mm and 8mm,
  box/.style={draw,rounded corners,align=center,minimum height=8mm,font=\footnotesize,fill=blue!5},
  rbox/.style={draw,rounded corners,align=center,minimum height=8mm,font=\footnotesize,fill=orange!10},
  gbox/.style={draw,rounded corners,align=center,minimum height=8mm,font=\footnotesize,fill=green!8},
  arr/.style={-{Latex[length=2mm]},thick}]
  \node[box] (wav) {Synthetic\\mixer / wav};
  \node[box,right=of wav] (fe) {log-mel\\ STFT};
  \node[rbox,below=8mm of fe] (rot) {Rotor\\estimator};
  \node[gbox,right=of fe] (bb) {Frozen\\foundation\\encoder\\(BEATs)};
  \node[rbox,right=of bb] (ad) {Rotor-aware\\adapter\\+ attn-pool};
  \node[box,right=of ad] (head) {HAP};
  \node[box,above=5mm of head,font=\scriptsize] (aux) {aux / domain\\(GRL)};
  \draw[arr] (wav)--(fe);
  \draw[arr] (wav) |- (rot);
  \draw[arr] (fe)--(bb);
  \draw[arr] (bb)--(ad);
  \draw[arr] (rot) -| (ad) node[midway,below,font=\scriptsize]{$r,\;R,\;M_{\text{res}}$};
  \draw[arr] (ad)--(head);
  \draw[arr] (ad)|-(aux);
\end{tikzpicture}}
\caption{EgoRAP pipeline. Rotor features $r$ (embedding), $R$ (harmonic mask) and
$M_{\text{res}}$ (residual mel) condition a lightweight adapter and attention pooling on top of
a frozen audio foundation model.}
\label{fig:framework}
\end{figure}
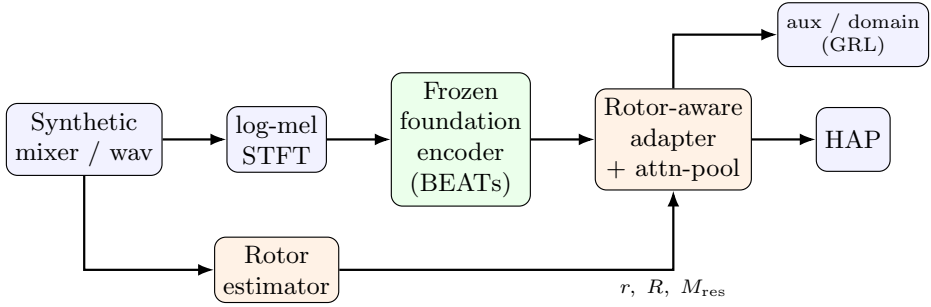

The compared models are frozen audio foundation encoders with small trainable heads. The most
capable configuration, EgoRAP-DA, augments a parameter-efficient BEATs adapter with two candidate
ingredients whose value is decided by real validation: conditioning on rotor harmonic structure
and domain-adversarial regularization against synthetic nuisance domains. These ingredients do
not yield a statistically significant gain on the limited real recordings, so they are presented
as part of a real-selected adapter family rather than as a standalone methodological claim.
Figure~\ref{fig:framework} summarizes the pipeline. A shared frontend produces a log-mel view, a
rotor estimator produces deterministic rotor features, a frozen foundation encoder produces token
embeddings, and the adapter head adapts and pools those embeddings for the HAP decision and
auxiliary/domain heads.

\subsection{Rotor harmonic estimation}
From the power spectrogram $|X(t,f)|^2$ we track the rotor blade-pass fundamental $f_0(t)$ in
$[80,1200]$~Hz with a batched harmonic-sum detector. The salience of a candidate $f_0$ is the
weighted sum of spectral magnitude at its first $K{=}8$ harmonics, and $f_0(t)$ is the temporal
median of the per-frame argmax. The soft rotor harmonic mask is
\begin{equation}
R(t,f)=\max_{k=1\ldots K}\exp\!\Big(-\frac{\big(f-k\,f_0(t)\big)^2}{2\sigma^2}\Big),
\end{equation}
and the rotor-suppressed residual spectrogram and mel representation are
\begin{equation}
M_{\text{res}}(t,f)=\log\!\big(\varepsilon+|X(t,f)|^2\,(1-R(t,f))\big).
\end{equation}
A fixed 14-dimensional descriptor $\rho$ summarizes statistics of $f_0$, harmonic-energy ratio,
voicing salience, low-band ratio, and residual flatness. A small MLP maps $\rho$ to a rotor
embedding $r$.

\begin{figure}[t]
\centering
\includegraphics[width=\textwidth]{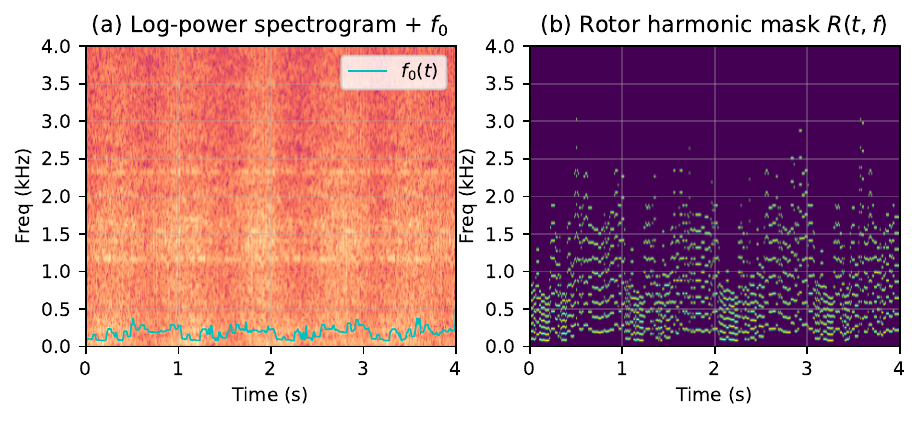}
\caption{Rotor estimation on a synthetic UAV clip. (a) Log-power spectrogram with the tracked
blade-pass fundamental $f_0(t)$. (b) The induced harmonic mask $R(t,f)$ concentrates on the rotor
comb, whose complement isolates broadband human cues.}
\label{fig:rotor}
\end{figure}

\subsection{Rotor-aware adaptation head}
Let $H\in\mathbb{R}^{N\times d}$ be the token sequence from the frozen encoder. Each rotor
adapter applies a bottleneck conditioned on $r$:
\begin{equation}
H' = H + W_{\text{up}}\,\phi\!\big(W_{\text{down}}\,\mathrm{LN}(H) + A\,r\big),
\end{equation}
with GELU $\phi$ and a zero-initialized $W_{\text{up}}$ so the adapter starts as an identity map.
A rotor-modulated attention pool aggregates tokens with a query $q+W_q r$:
\begin{equation}
\alpha = \mathrm{softmax}\!\Big(\tfrac{H'(q+W_q r)}{\sqrt{d}}\Big),\qquad
z = \textstyle\sum_n \alpha_n H'_n .
\end{equation}
The pooled vector $z$ feeds a HAP head, an auxiliary four-way source head, a
supervised-contrastive projection, and, through a gradient-reversal layer, drone and acoustic
domain classifiers.

\subsection{Training objective and real-validated configuration}
All variants share the base objective
\begin{equation}
\mathcal{L}=\mathcal{L}_{\text{focal}}
+\lambda_{\text{aux}}\mathcal{L}_{\text{aux}}
+\lambda_{\text{cons}}\mathcal{L}_{\text{cons}}
+\lambda_{\text{dom}}\mathcal{L}_{\text{grl}}
+\lambda_{\text{sc}}\mathcal{L}_{\text{supcon}} ,
\end{equation}
where $\mathcal{L}_{\text{focal}}$ is focal loss on HAP, $\mathcal{L}_{\text{aux}}$ is
cross-entropy on the auxiliary source class, $\mathcal{L}_{\text{cons}}$ ties HAP probability to
the auxiliary human-source probability, $\mathcal{L}_{\text{grl}}$ is a domain-adversarial term
applied through a ramped gradient-reversal layer, and $\mathcal{L}_{\text{supcon}}$ clusters
same-HAP samples across nuisance domains. Only the rotor MLP, adapters, pooling, and heads are
trained; the backbone stays frozen.

The configuration is fixed by real validation rather than by synthetic accuracy. We treat rotor
conditioning, domain-adversarial strength, and supervised contrastive learning as a small
candidate set and select among them only on Real-Dev operating data. The selected configuration
uses rotor conditioning, moderate domain-adversarial regularization, and the contrastive term; we
call this Real-Dev-selected configuration EgoRAP-DA.

\begin{algorithm}[t]
\caption{EgoRAP training step}
\label{alg:egorap}
\begin{algorithmic}[1]
\Require frozen encoder $g$; frontend; rotor estimator; head $h_\theta$
\State sample batch $\{x_i,y_i,c_i,d_i\}$ from online or cached synthetic mixtures
\State $(\text{logmel}, |X|^2) \gets \text{frontend}(x)$ \Comment{fp32}
\State $\rho, R, M_{\text{res}} \gets \text{RotorEstimator}(|X|^2)$
\State $H \gets g(\text{logmel/wav})$ \Comment{frozen, no grad}
\State $\{o_{\text{hap}},o_{\text{aux}},z,o_{\text{dom}}\} \gets h_\theta(H,\rho;\lambda_{\text{grl}})$
\State $\mathcal{L}\gets$ focal $+\lambda_{\text{aux}}$aux$+\lambda_{\text{sc}}$supcon$+\lambda_{\text{dom}}$grl$+\lambda_{\text{cons}}$cons
\State update $\theta$ with AdamW and AMP; keep backbone unchanged
\end{algorithmic}
\end{algorithm}

\section{Experimental Setup}\label{sec:setup}
\subsection{Backbones and candidates}
The cross-architecture baselines are SE-ResNet18 trained from scratch on log-mel features, PANNs
CNN14 with a frozen embedding, AST pretrained on AudioSet and frozen, and BEATs
(\texttt{iter3+} SSL) frozen. On top of frozen BEATs we compare a linear probe, a plain adapter
without rotor or domain loss, a rotor-conditioned adapter, domain-adversarial variants with
$\lambda_{\text{dom}}\in\{0.05,0.3\}$, a rotor+DA+SupCon variant, and a SpecAugment ablation.
EgoRAP uses BEATs by default, and weaker-backbone swaps are reported only as ablations.

The final method is the candidate that maximizes Real-Dev operating recall at $5\%$ FPR subject to
AUROC being within $0.03$ of the best Real-Dev AUROC. Ties are broken by recall at $1\%$ FPR and
then by EER. This rule fixes EgoRAP-DA before Real-Test is scored.

\subsection{Training and metrics}
Training uses 4~s clips at 16~kHz, batch size $64$, AdamW with learning rate
$3\!\times\!10^{-4}$ and weight decay $10^{-4}$, a one-cycle schedule, $14$ epochs, AMP, and
fp32 execution for frozen foundation backbones. Base loss weights are
$\lambda_{\text{aux}}{=}0.3$, $\lambda_{\text{sc}}{=}0.3$, and
$\lambda_{\text{cons}}{=}0.1$; $\lambda_{\text{dom}}$ is selected as part of the candidate grid.

The primary metric is recall at $5\%$ FPR, which reflects the low-false-alarm operating point needed
by an SAR alarm. AUROC, AUPRC, recall at $1\%$ FPR, and EER are reported alongside. Real intervals
are $95\%$ group bootstraps over recordings with $2000$ resamples, synthetic evaluation sets are
fixed by seed, and the real test is scored only after Real-Dev selection is complete.

\subsection{Hardware and software}
Experiments were run on a Linux 5.15 system with four Tesla V100-DGXS-32GB GPUs and NVIDIA driver
535.230.02. The software environment used Python 3.13.12, PyTorch 2.6.0+cu124, torchaudio
2.6.0+cu124, CUDA 12.4, NumPy 2.4.6, scikit-learn 1.9.0, and Matplotlib 3.10.9. Single-clip
latency in Table~\ref{tab:eff} was measured on a V100 GPU.

\section{Results}\label{sec:results}
\begin{table}[t]
\centering
\caption{Synthetic vs.\ locked real results. Synthetic column is in-distribution AUROC (M0); real columns are the operating regime ($\tau{=}75$~dB, 1711 audible positives / 58 recordings), with $95\%$ \emph{group} bootstrap CIs (resampling recordings). The primary metric is recall at $5\%$ FPR. The main method ($\star$) was fixed on Real-Dev (Table~\ref{tab:realdev}) before these numbers were computed; no real-test number was used for selection. Over 3 seeds the selected model attains real AUROC 0.744$\pm$0.026 and recall at $5\%$ FPR 0.549$\pm$0.014.}
\label{tab:realtest}
\resizebox{\textwidth}{!}{%
\begin{tabular}{lccccc}
\toprule
Model & Synth AUROC & Real AUROC [95\% CI] & Real R@FPR5 [95\% CI] & R@FPR1 & EER \\
\midrule
SE-ResNet18 & 0.743 & 0.536 [0.394,0.682] & 0.148 [0.060,0.260] & 0.108 & 0.484 \\
AST & 0.649 & 0.721 [0.584,0.840] & 0.176 [0.057,0.542] & 0.061 & 0.318 \\
BEATs (probe) & 0.741 & 0.755 [0.627,0.854] & 0.504 [0.338,0.659] & 0.468 & 0.334 \\
BEATs-Adapter & 0.831 & 0.763 [0.635,0.863] & 0.483 [0.335,0.637] & 0.444 & 0.314 \\
EgoRAP (no DA) & 0.831 & 0.776 [0.652,0.872] & 0.508 [0.361,0.673] & 0.484 & 0.301 \\
EgoRAP-DA + SupCon\,$\star$ & 0.751 & 0.753 [0.642,0.845] & 0.559 [0.405,0.689] & 0.525 & 0.302 \\
\bottomrule
\end{tabular}}
\end{table}

\subsection{Synthetic scores misrank real transfer}
\begin{figure}[t]
\centering
\includegraphics[width=\textwidth]{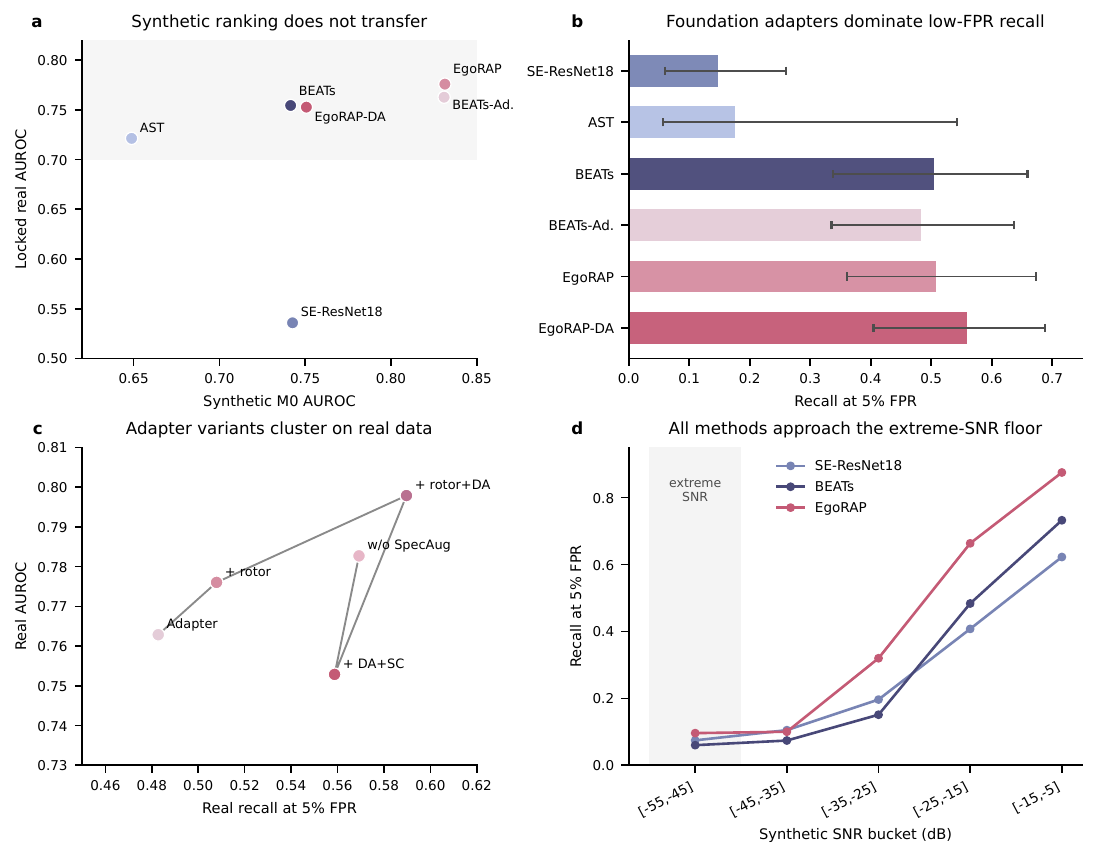}
\caption{Nature-style summary of the real-validation evidence. (a) Synthetic in-distribution
AUROC weakly and non-monotonically tracks locked real-operating AUROC. (b) Locked-test
recall at $5\%$ FPR with group-bootstrap intervals shows that foundation models and adapters are far
stronger than a from-scratch CNN under real ego-noise. (c) Adapter-family ablations cluster
closely, so component-level rankings remain uncertain. (d) Synthetic SNR buckets show the
detection envelope and the collapse below $-45$~dB.}
\label{fig:naturesummary}
\end{figure}

Table~\ref{tab:realtest} and Fig.~\ref{fig:naturesummary} show that synthetic AUROC is not a
reliable proxy for real UAV transfer. SE-ResNet18 is competitive on synthetic M0 AUROC
($0.743$) but falls to real AUROC $0.536$ and recall at $5\%$ FPR $0.148$. AST shows the opposite
ordering, with lower synthetic AUROC ($0.649$) but substantially stronger real AUROC ($0.721$).
Frozen foundation encoders and their adapters all transfer to roughly $0.72$--$0.79$ real AUROC.
The ranking reversal means that selecting by synthetic accuracy can choose a poor real operating
model.

\begin{table}[t]
\centering
\caption{Pre-registered model selection on Real-Dev (operating, $\tau{=}75$~dB, 1109 audible positives / 38 recordings). Selection rule: maximize recall at $5\%$ FPR subject to AUROC$\ge$best$-0.03$, tie-break by recall at $1\%$ FPR and then EER. Selection uses \emph{only} this split; the reported test numbers (Table~\ref{tab:realtest}) are never used to choose the method.}
\label{tab:realdev}
\begin{tabular}{lccccc}
\toprule
Configuration & AUROC & R@FPR5 & R@FPR1 & EER & sel. \\
\midrule
BEATs-Adapter & 0.760 & 0.417 & 0.383 & 0.303 & $\checkmark$ \\
EgoRAP (no DA) & 0.752 & 0.499 & 0.471 & 0.326 &  \\
EgoRAP-DA (light) & 0.785 & 0.568 & 0.543 & 0.298 & $\checkmark$ \\
EgoRAP-DA (strong) & 0.786 & 0.567 & 0.473 & 0.287 & $\checkmark$ \\
EgoRAP-DA + SupCon\,$\star$ & 0.773 & 0.592 & 0.527 & 0.300 & $\checkmark$ \\
EgoRAP-DA w/o specaug & 0.747 & 0.567 & 0.535 & 0.326 &  \\
\bottomrule
\end{tabular}
\end{table}

\subsection{Real-Dev selection fixes the reported method}
The method is selected only on Real-Dev operating data using the pre-registered rule in
Table~\ref{tab:realdev}. Four adapter configurations satisfy the AUROC constraint, and the
rotor+DA+SupCon configuration has the highest Real-Dev recall at $5\%$ FPR ($0.592$). This configuration
is fixed as EgoRAP-DA before the locked Real-Test is evaluated.

\subsection{Locked real-test performance}
On Real-Test, EgoRAP-DA gives the best low-false-alarm recall in the main comparison
(recall at $5\%$ FPR $0.559$ versus $0.483$ for BEATs-Adapter), while its AUROC ($0.753$) is statistically
indistinguishable from the adapter's ($0.763$). Under paired group bootstrap, the recall
difference is $+0.069$ with $95\%$ CI $[-0.017,+0.133]$; the gain is
positive in $0.956$ of bootstrap resamples but not significant at the $95\%$ level. The honest
reading is that the adapter family is better suited to real UAV audio than from-scratch CNNs, but
the limited number of operating recordings does not resolve fine-grained ranking among adapter
variants.

\begin{table}[t]\centering
\caption{Recall at $5\%$ FPR by SNR bucket (M0). EgoRAP's advantage concentrates at mid/low SNR.}
\label{tab:snr}
\resizebox{\columnwidth}{!}{%
\begin{tabular}{lccccc}
\toprule
Model & $[-55,-45]$ & $[-45,-35]$ & $[-35,-25]$ & $[-25,-15]$ & $[-15,-5]$ \\
\midrule
SE-ResNet18 & 0.075 & 0.105 & 0.196 & 0.408 & 0.622 \\
BEATs & 0.060 & 0.074 & 0.151 & 0.484 & 0.732 \\
EgoRAP-BEATs (ours) & 0.096 & 0.100 & 0.320 & 0.663 & 0.875 \\
\bottomrule
\end{tabular}}
\end{table}

\begin{table}[t]\centering
\caption{Efficiency: parameters, trainable parameters, and single-clip GPU latency (V100). RTF $<1$ is faster than real time.}
\label{tab:eff}
\begin{tabular}{lcccc}
\toprule
Model & Params (M) & Train (M) & ms/4s & RTF \\
\midrule
SE-ResNet18 & 11.6 & 11.64 & 9.8 & 0.002 \\
PANNs CNN14 & 86.5 & 4.62 & 10.6 & 0.003 \\
AST & 86.9 & 0.76 & 40.0 & 0.010 \\
BEATs & 91.1 & 0.76 & 51.5 & 0.013 \\
BEATs-Adapter & 91.3 & 0.97 & 50.4 & 0.013 \\
EgoRAP-BEATs & 91.3 & 0.97 & 50.9 & 0.013 \\
\bottomrule
\end{tabular}
\end{table}

\subsection{Detection envelope and efficiency}
The synthetic SNR analysis in Table~\ref{tab:snr} and Fig.~\ref{fig:analysis} places the real
results in a detection envelope. EgoRAP's gain concentrates at low and mid SNR, but all methods
degrade sharply below $-45$~dB, where positives approach the physical detection floor. Rotor
conditioning and domain regularization add negligible parameters and latency over the plain
adapter, so the real-test comparison is between models of essentially equal runtime cost.

\section{Ablation and Analysis}\label{sec:ablation}
\begin{table}[t]
\centering
\caption{Component ablation on the locked real test (operating). Adapter-family variants cluster within group-bootstrap noise on real AUROC; the pre-registered main method ($\star$) has the best low-false-alarm recall. The robust lift is foundation adaptation, not any single add-on.}
\label{tab:ablation}
\begin{tabular}{lccc}
\toprule
Configuration & Synth AUROC & Real AUROC & Real R@FPR5 \\
\midrule
BEATs-Adapter (no rotor/DA/SC) & 0.831 & 0.763 & 0.483 \\
+ rotor conditioning & 0.831 & 0.776 & 0.508 \\
+ rotor + DA ($\lambda{=}0.3$) & 0.739 & 0.798 & 0.590 \\
+ rotor + DA + SupCon ($\star$) & 0.751 & 0.753 & 0.559 \\
\;\; ablate SpecAugment & 0.756 & 0.783 & 0.569 \\
\bottomrule
\end{tabular}
\end{table}

Table~\ref{tab:ablation} adds the adapter ingredients one at a time on the locked real test. The
plain BEATs adapter already provides most of the robust transfer gain, and rotor conditioning
nudges both real AUROC and low-FPR recall upward. Domain-adversarial training lowers synthetic
AUROC while improving real recall, which illustrates why synthetic validation alone can discard a
useful component. The variants nevertheless remain within broad group-bootstrap uncertainty on
real data, and an unselected strong-DA variant reaches the highest real AUROC and recall at
$5\%$ FPR.
This pattern supports the selection protocol but does not support over-reading the exact ordering
among adapter variants.

\begin{figure}[t]
\centering
\begin{minipage}{0.49\textwidth}
\centering
\includegraphics[width=\linewidth]{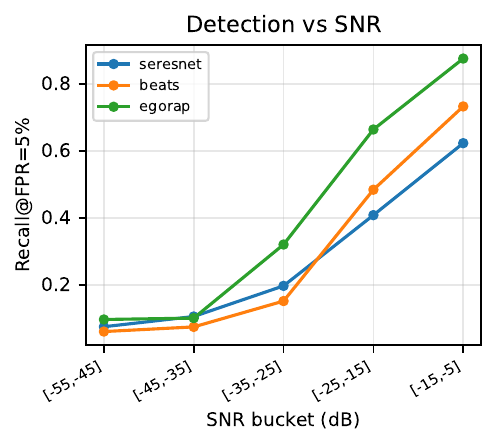}\\[-1mm]
{\small (a)}
\end{minipage}
\hfill
\begin{minipage}{0.49\textwidth}
\centering
\includegraphics[width=\linewidth]{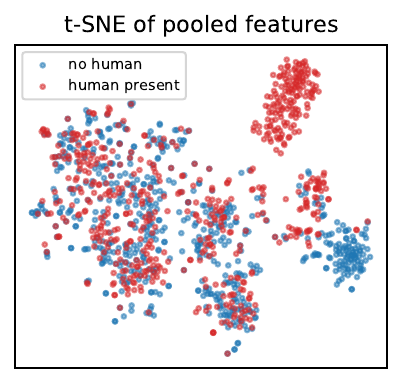}\\[-1mm]
{\small (b)}
\end{minipage}
\caption{Additional analysis. (a) Recall at $5\%$ FPR by synthetic SNR bucket. (b) t-SNE of pooled
features on source-OOD M3, colored by human presence.}
\label{fig:analysis}
\end{figure}

The qualitative diagnostics are consistent with the quantitative results. Fig.~\ref{fig:rotor}
shows the estimated rotor comb aligned with blade-pass harmonics, making the residual view
emphasize broadband human energy. Fig.~\ref{fig:analysis} shows both the SNR envelope and the
pooled-feature separation on source-OOD data. Failure cases concentrate on very distant or brief
vocalizations whose energy is genuinely below the rotor floor.

\section{Discussion and Limitations}\label{sec:discussion}
The central takeaway is that group-level real validation makes apparent method gains smaller and
less certain, while leaving two conclusions firm: synthetic accuracy is a poor proxy for real UAV
transfer, and foundation adapters transfer far better than synthetic-trained CNNs. EgoRAP-DA is
therefore presented as the configuration selected by a disciplined real-validation procedure, not
as a statistically significant new state of the art over a vanilla adapter.

The operating regime also clarifies what remains unsolved. Reporting a single score over all
positives is misleading when many positives are physically inaudible, so the evaluable regime is
defined from metadata alone. Even inside that regime, the best model still misses over $40\%$ of
audible positives at a $5\%$ false-alarm budget, and below $-45$~dB all methods sit near the
detection floor. These numbers establish a realistic baseline rather than a deployable SAR
system.

The main limitation is the size and diversity of the real operating set. It contains only dozens
of operating recordings from a single corpus, so group-bootstrap intervals are wide, method
differences among adapter variants are not significant, and cross-corpus generalization remains
untested. The current study also uses a single channel, although DroneAudioSet arrays and rotor
phase could support spatial rotor cancellation. Finally, the synthetic ego-noise pool covers a
limited platform set; broadening it may improve transfer, but any such improvement needs to be
confirmed on real grouped data.

\section{Conclusion}\label{sec:conclusion}
Synthetic mixtures alone do not provide a dependable basis for judging UAV audition models.
Synthetic accuracy is a poor, non-monotonic proxy for real transfer, badly enough to reorder
models and make a from-scratch CNN look competitive while it collapses on real ego-noise. The
main contribution is a real-validated benchmark and protocol with operating-level audibility,
recording-grouped Dev/Test splits, and group-bootstrap intervals. Within this protocol, a
domain-regularized BEATs adapter selected on Real-Dev gives the best low-false-alarm operating
point on the locked test, though its advantage over a vanilla adapter is not significant given the
available real recordings. The dependable message is that foundation adaptation transfers, and
that real, group-level validation is necessary for honest progress in UAV audition.

\section{Acknowledgement}
This work (MPU submission code: fca.1db7.b77d.f) is supported by the grant from Macao Polytechnic University (RP/FCA-06/2026) and the Macao Science and Technology Development Fund (FDCT-MOST: 0018/2025/AMJ).

\bibliographystyle{splncs04}
\bibliography{refs}

\end{document}